\documentclass{ws-ijmpa_mod}

\newcommand{\minerva}{\mbox{\hbox{MINER}$\nu$\hbox{A}}}
\newcommand{\numi}{\mbox{NuMI}}
\newcommand{\minos}{\mbox{MINOS}}

\begin{document}


%
\catchline{}{}{}{}{}
%

\title{The Design and Performance of the \minerva\ Detector }

\author{\footnotesize Howard Budd}

\address{Department of Physics and Astronomy, 
 The University of Rochester\\
Rochester, New York 14627-0171, USA}

\maketitle


\begin{abstract}
The \minerva\ experiment is designed to make precision measurements
of various neutrino cross sections in the low energy regime.
We describe the detector and give the performance of 
some of the measured quantities.

\end{abstract}

\section{The Goals of the \minerva\ Detector}
  The \minerva\ detector~\cite{minerva} is a fine-grain 
neutrino detector, which sits in the \numi\
beam line. The \minerva\ detector must be able to reconstruct 
exclusive final states. This requires high granularity for 
charged particle tracking, particle ID, and low momentum thresholds.
The detector must contain both electromagnetic 
and hadronic showers 
and  measure muon momentum. 
Hence, the fiducial volume of the detector should be 
fine grained, fully active, and surrounded by calorimeters.
To measure nuclear effects, the detector should contain
targets of different nuclei. 

\section{The \minerva\ Detector}

\begin{figure}
\vspace*{-2ex}
\begin{center}
\epsfxsize=0.513\textwidth\epsfbox{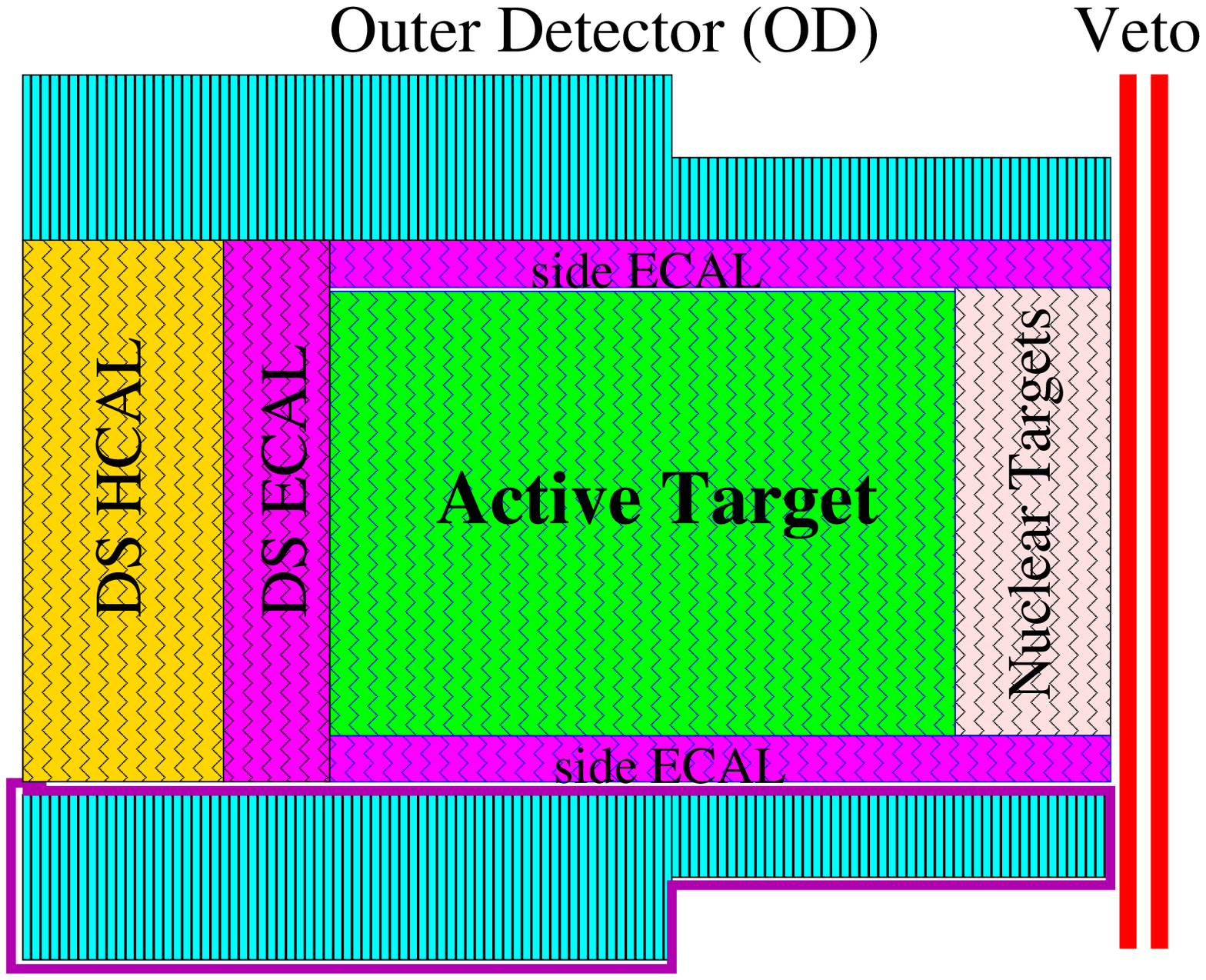}
\hfil
\epsfxsize=0.36\textwidth\epsfbox{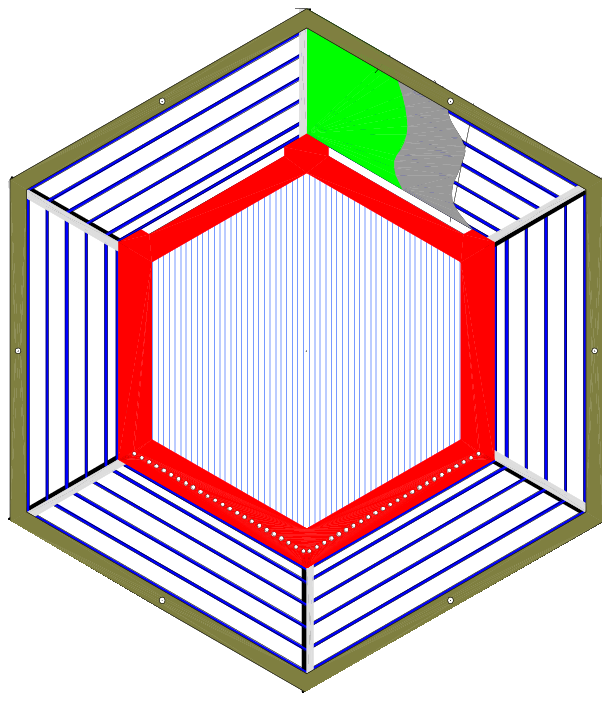}
\end{center}
\caption[The \minerva\ detector]{Left: A schematic side
  view of the full \minerva\ detector with sub-detectors labeled.  The
  neutrino beam enters from the right.  Right: A slice through the active target
  region of the detector, showing the Inner Detector (blue), side ECAL (red) and
  surrounding outer detector. Right: A beam's eye view of the detector showing
the active target and side calorimeters.}
\label{fig:detector}
\end{figure}

Figure~\ref{fig:detector} shows a schematic view of the \minerva\ detector.
The various sub-detectors include 
the active target, the nuclear targets, the downstream and side ECAL, 
the downstream and side HCAL, and the veto. 
The detector is 3.88 m  across, 4.48 m high, and 4.7 m long.

The inner active target consists of fully active scintillating strips  
read out by wave length shifting (WLS) fibers. The active target 
has 6 tons of plastic with 3-5 tons of fiducial 
volume, depending upon the physics process. The scintillating strips
are extruded triangular strips with a 3.3~cm base and 1.7~cm height. The
strips are triangular to improve the coordinate resolution by
using light sharing between adjacent strips. 
The active target strips are organized in   
modules which have 4 planes. Each plane is shaped as a hexagon
with 128 strips. The hexagon shape enables
the planes to have orientations which differ by 60$^\circ$.
The orientation of the planes for a module is X, U, X, V. The
X layers seed the track reconstruction, and the U V stereo layers 
identity and reconstruct the three dimensional tracks.

The downstream and side calorimeters 
yield the calorimetic energy 
of the events. These sampling detectors use the same
scintillating strips as the active target.
The side ECAL has 2 mm lead
plates mounted on the frames and extend 15 cm into
the central scintillator region, see Figure~\ref{fig:detector}.
The side HCAL acts as a frame for the active target.
The side HCAL has 6 scintillator layers with
50 cm of steel between the active target and the outer 
most scintillator strip. The upstream parts of the side HCAL has 4 active 
elements with a total of 30 cm of steel 
between the active target and the outer
most scintillator strip.
The downstream ECAL 
calorimeter has  20 planes of scintillator  
with 5/16"  lead plates attached in front of each plane.
The downstream HCAL consists of 20 planes of 
scintillator with 1" thick steel plates in front of each plane.
The bottom of the right side of  Figure~\ref{fig:detector} shows holes
for the magnetic coil. The  coils provide a 
field of 1.6 Tesla in the side HCAL.

The upstream nuclear targets has 3 plates each of 
graphite, steel, and lead, for a total of
9 plates of nuclear targets.  The nuclear targets have 1/2, 1, and 
1 fiducial tons of graphite, steel, and lead, respectively.
The graphite target (pure carbon)  provides consistency check
for the scintillator target (carbon and hydrogen). 
Between each plate are 4 planes of scintillator.
The nuclear targets section also serves as the upstream calorimeters.

Hamamatsu R7600U-00-M64 PMTs (64 channel multi-anode 
photo multiplier tubes) collect the scintillation
light from the optical fibers. These tubes are very 
similar to the PMTs used by the \minos\
Near Detector. The PMTs provide both ADC and TDC information 
and sit in optical boxes on top of the detector.

\minerva\ uses the  D0 TRiP ASIC~\cite{rubinov:trip} to readout the PMTs.  The
TRiP chip is a redesign of the readout ASIC for the D0 fiber tracker
and preshower. The analogue portion of TRiP chip is based
on the SVX4 chip and can store 4 analogue buffers to be readout. 
Each PMT channel is connected to 2 TRiP channels providing
both a low gain and high gain ADC. Since a TRiP chip has 32 channels, 
each chip can service 16 PMT channels with both ADC and TDC information. 
The TDC information comes from a latch in the 
TRiP chip  which is fed into a 
FPGA. The timing information has an accuracy of a few ns.
When this latch fires, the FPGA tells the TRiP chip 
to store the analogue charge. 
This enables a TRiP chip to 
store as many as 4 events throughout the 8 $\mu$ sec spill.
At the end of the spill, the VME readout collects all the 
TRiP chip information from the FPGAs. 

 The TRiP electronics has been tested with a vertical slice test.
We have built a readout board which 
plugs on the \minos\ CalDet box~\cite{Tagg:pmt}. 
The CalDet box contains a 64 channel PMT and optical fibers
to bring the light to the PMT. (These boxes were used for
the \minos\ CERN test beam). The pedestal RMS 
is about 2 fc, much less than the 30 fc photo electron peak 
we expect from the least gain PMT channel.  We have verified the 
self-triggering mode for storing charge. We have determined
that 2 TRiP channels per PMT channel gives the dynamic range we need.  

The \minerva\ detector sits just upstream of the \minos\
detector in the \minos\ near detector hall. The \minos\
detector  serves to measure the charge and 
momentum of the higher energy muons exiting the \minerva\ detector.

\section{The Performance of the \minerva\ detector}

\begin{figure}
\centerline{
\epsfig{figure=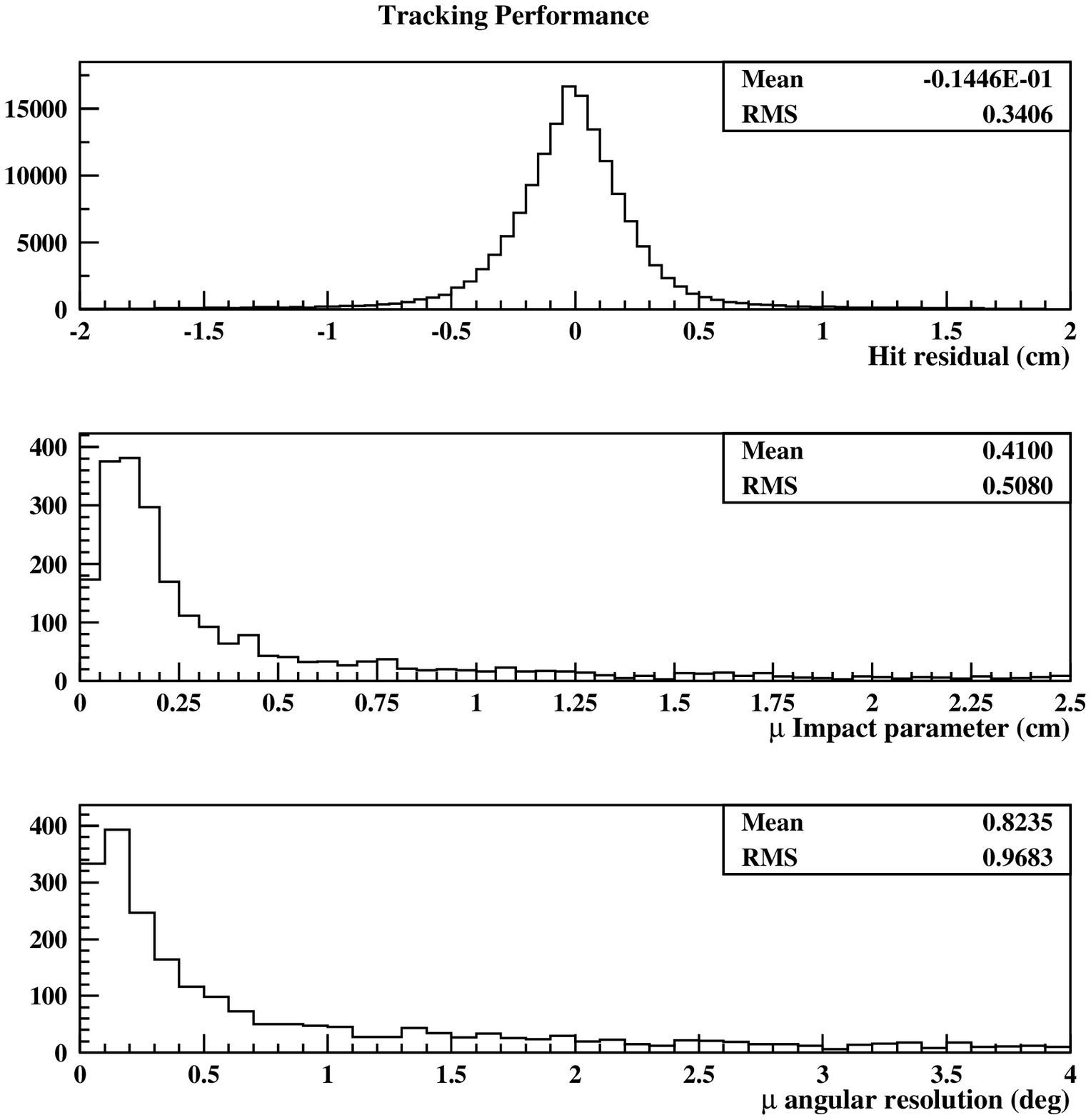,width=0.47\textwidth,bbllx=50,bblly=150,bburx=555,bbury=669}
\hfil
\epsfig{figure=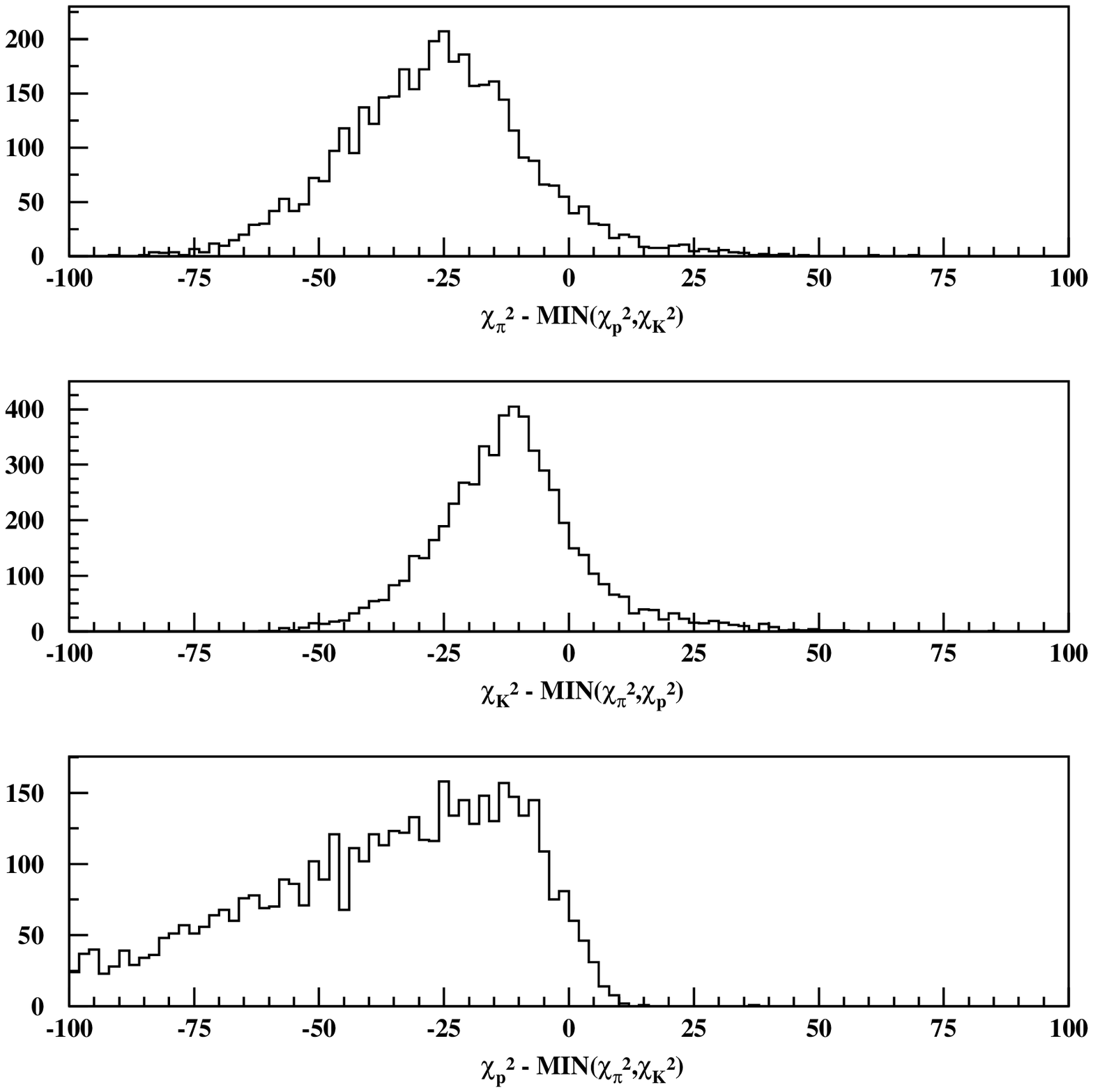,width=0.52\textwidth,bbllx=0,bblly=13,bburx=567,bbury=540}}\vspace*{-0.2in}
\caption[Tracking and particle identification performance.]{Left: Coordinate resolution, impact parameter and
angular resolution for muons produced in CC reactions.
Right: the $\Delta\chi^2$~$dE/dx$~estimator for simulated charged pions(top),
kaons(middle) and protons(bottom) stopping in the inner detector.
Tracks with $\Delta\chi^2 < 0$ are correctly identified.}
\label{fig:recon}
\end{figure}

The simulation of events in the \minerva\
detector is carried out using a GEANT-based Monte Carlo 
program with the \numi\ beam flux. The Monte Carlo
program includes the photostatistical effects of
light collection, a Kalman filter reconstruction package
for track and vertex fitting, and particle identification.
Some of the results of the simulation are shown in
Figure~\ref{fig:recon}. Fitted tracks from such 
muons have typical impact parameter of $\sim 2$~mm 
and angular resolution of $<9$~mrad (Figure~\ref{fig:recon}).
Using the (typically short) reconstructed proton track and the muon
track from quasi-elastic events, RMS vertex
uncertainties of $9$~mm and $12$~mm are measured in the coordinates
transverse and parallel to the beam direction, respectively.

 Specific energy loss ($dE/dx$) will be an important tool 
for particle identification in \minerva.
 For tracks which stop in the inner detector, the charge 
deposited near the end of the track (corrected for sample length) 
can be compared with expected curves for 
the $\pi^\pm$, $K^\pm$ and proton hypotheses.
Figure~\ref{fig:recon} illustrates the
probability of misidentification by plotting the difference
$\Delta\chi^2$ between the correct $\chi^2$ (for the particle's true
type) and the smallest of the two (incorrect) other particle
hypotheses.  With this naive $dE/dx$ analysis,
we correctly identify 85\% of stopping kaons, 90\% of stopping pions, and
$>95\%$ of stopping protons.

\begin{figure}
\begin{center}
\begin{minipage}[b]{0.49\textwidth}
\epsfxsize=1.\textwidth\epsfbox{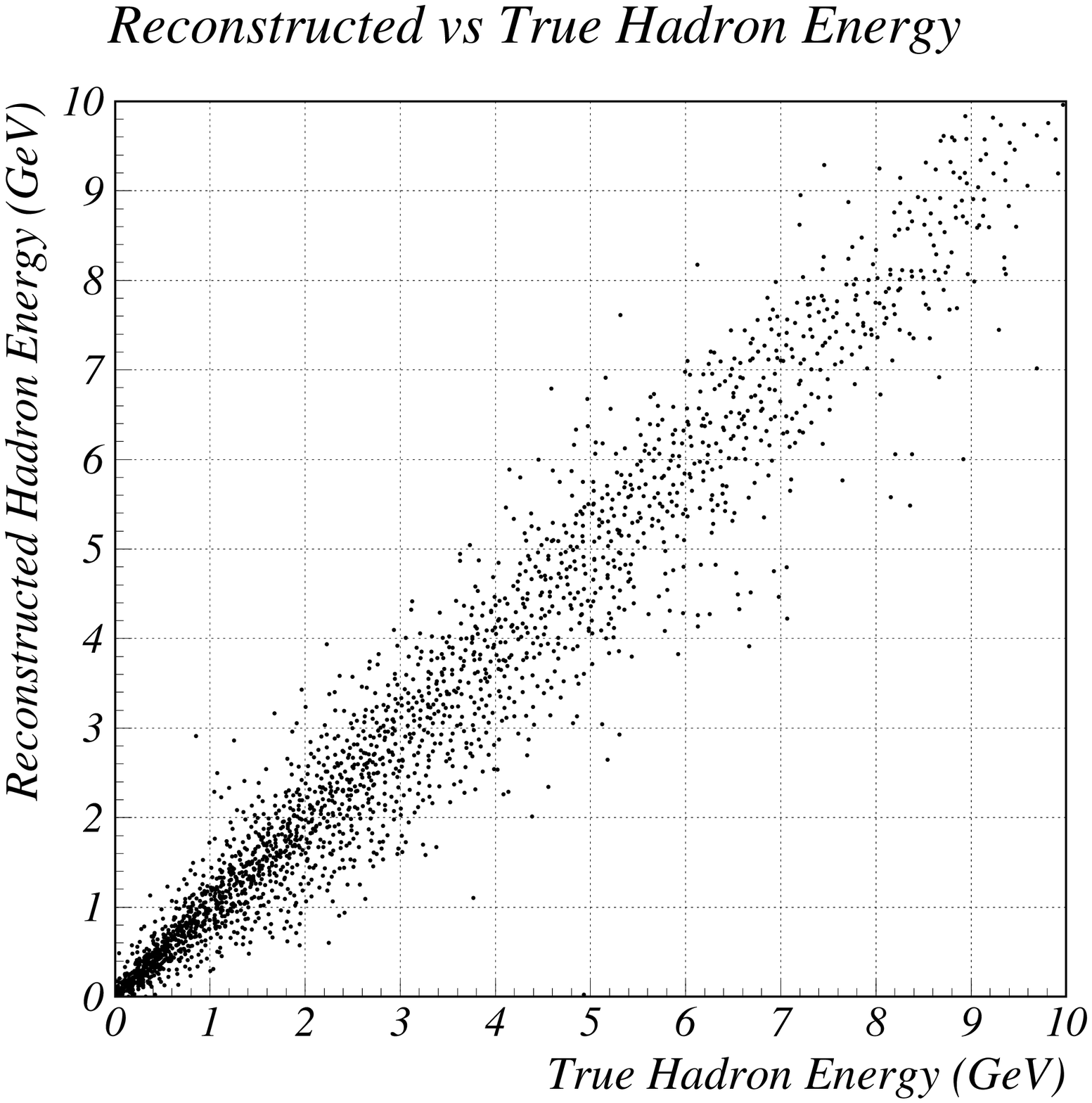}
\end{minipage}
\begin{minipage}[b]{0.49\textwidth}
\epsfxsize=1.\textwidth\epsfbox{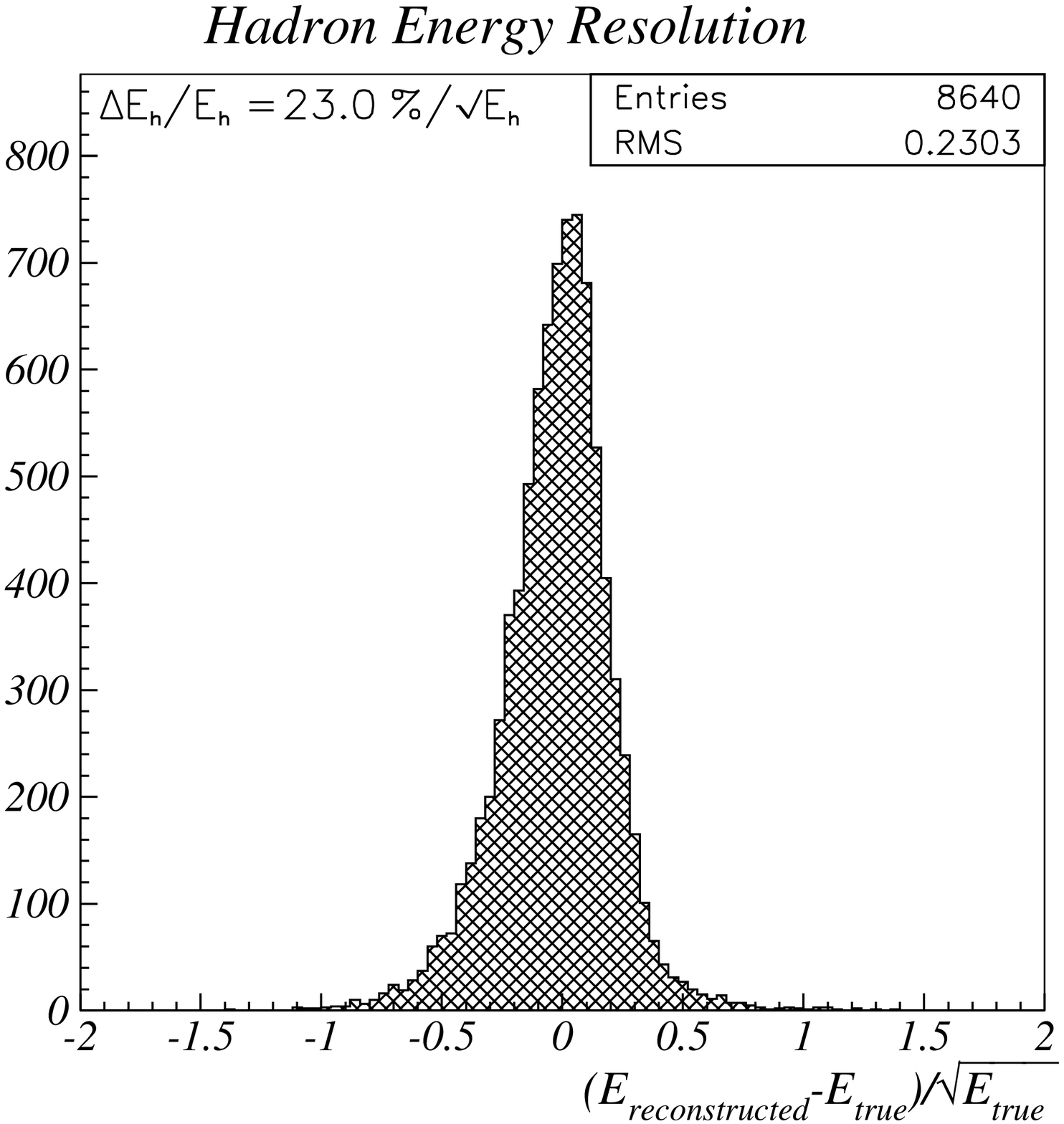}
\end{minipage}\vspace*{-0.1in}
\end{center}
\caption[Fit and resolution of $E_h$ determination]{
The left graph shows on the vertical axis the the hadronic energy
$E_h$ reconstructed from scintillator output in {\minerva} vs. the
true $E_h = E_{\nu} - E_{\mu}$.   Right Figure shows the relative deviation
of the fit, $(\Delta E_h/E_h) \sqrt{E_h}$ vs. the true $E_h$.}
\label{res:ehfit}
\end{figure}


The energy of muons from charged-current interactions will be measured using
range and/or curvature in the magnetized regions of \minerva\ and the MINOS
spectrometer. For muons
stopping in the detector, the momentum resolution will
be $\frac{\Delta p}{p} \sim 5\%$. If the MINOS detector is
used, the momentum resolution will be 13\%\cite{MINOS_TRD_NDHALL}.

Figure~\ref{res:ehfit} shows the reconstructed $E_h$  vs the
true $E_h$ for  Monte Carlo events where all of
the hadronic fragments are contained within the {\minerva} detector.
The relative deviation of the reconstructed energy from the true
$E_h$, $\Delta E_h / E_h$, multiplied by $\sqrt{E_h}$ is shown in
Figure~\ref{res:ehfit}, giving an average resolution for reconstruction of
$E_h$ of ${\Delta E_h}/{E_h} = {23\%}/{\sqrt{E_h(\rm GeV)}}$.
This $1/{\sqrt{E_h}}$ resolution has some energy dependence and is best
represented by
${\Delta E_h}/{E_h} = 4\% + {18\%}/{\sqrt{E_h(\rm GeV)}}$.


\begin{thebibliography}{0}
\bibitem{minerva} \minerva\ Collaboration, Proposal for
Fermilab Experiment E938, hep-ex[0405002]

\bibitem{rubinov:trip} 
J. Estrada, C. Garcia, B. Hoeneisen and P. Rubinov, Aug 2002,
FERMILAB-TM-2226.

\bibitem{Tagg:pmt} 
N. Tagg et al., submitted to Nucl. Instr. and Meth., physics[0408055]

\bibitem{MINOS_TRD_NDHALL} 
MINOS Collab., ``MINOS Technical Design Report''
NuMI-NOTE-GEN-0337 (1998).

\end{thebibliography}
\end{document}